\documentclass[12pt a4paper]{article}

\usepackage{latexsym}
\usepackage{graphicx}
\usepackage{amsmath}
\usepackage{amssymb}
\graphicspath{{./}{./figs/}}

\begin{document}

\title{\bf Controlling the spreading in small-world networks}
\author{{\bf Xiang Li$\,^1$\thanks{xli@sjtu.edu.cn},
\bf Guanrong Chen$\,^2$\thanks{eegchen@cityu.edu.hk}}\\
\\
$\,^1$ Dept. of Automation, Shanghai Jiaotong University\\
Shanghai, 200030, P.R. China\\
\\
$\,^2$ Dept. of Electronic Engineering, City University of Hong Kong\\
Hong Kong, P.R. China\\} \maketitle

{\centering (The 23rd Chinese Control Conference at Wuxi, China,
August, 2004)\\}

\begin{abstract}

The spreading (propagation) of diseases, viruses, and disasters
such as power blackout through a huge-scale and complex network is
one of the most concerned issues today. In this paper, we study
the control of such spreading in a nonlinear spreading model of
small-world networks. We found that the short-cut adding
probability $p$ in the N-W model \cite{N-W:1999} of small-world
networks determines the Hopf bifurcation and other bifurcating
behaviors in the proposed model. We further show a control
technique that stabilize a periodic spreading behavior onto a
stable equilibrium over the proposed model of small-world networks.\\


\noindent{Correspondence author:}\\
Dr. Xiang Li\\
Dept. of Automation\\
Shanghai Jiaotong University\\
Shanghai 200030, P.R. China\\
Telephone: 86-21-62831590\\

\end{abstract}

\section{Introduction}

\noindent{R}ecently, Watts and Strogatz proposed a small-world
model \cite{W-S:1998}, named W-S model, which consists of a
rewired regular lattice with a very small fraction of long-range
connections. Newman and Watts \cite{N-W:1999} then made a slight
modification on the rewiring process of the W-S model, named N-W
model, by means of adding linkages between pairs of randomly
chosen nodes with a very small probability $0<p\ll 1$. When
selecting a very small probability $0<p\ll 1$ in these models, the
obtained small-world networks have large clustering and small
average distances, which match the familiar small-world feature
discovered in many real-life large-scale networks
\cite{L-J-C:2003,Newman:2000,S-B:2003,W-C:2003,Watts:1999}.\\
\indent{Since} the discovery of the small-world phenomenon, many
interesting results on the analysis, control and applications of
small-world networks have been published \cite{Mou:1999, W-C:2003,
Watts:1999, Yang:2001}. In particular, Moukarzel \cite{Mou:1999}
studied a linear disease spreading model of small-world networks,
which was further extended to a nonlinear model with frictions and
time-delays where chaos and bifurcation can emerge
\cite{L-C-L:2004,Yang:2001}. In this paper, we concern with the
Hopf bifurcation phenomenon during the spreading process in
small-world networks, and its control problem, by utilizing the
short-cut adding probability $p$ as the means to stabilize the
bifurcating behaviors. \\
\indent{The} remainder of this paper is organized as follows:
Section 2 describes the nonlinear spreading model, where the
existence of Hopf bifurcation in this model is discussed. Section
3 proposes a method of of controlling the bifurcating behaviors
using the short-cut adding probability $p$. The effectiveness of
this control technique is visualized via a numerical example, also
in this section. Finally, Section 4 concludes the paper.

\section{Nonlinear spreading model and Hopf bifurcation}

\indent{Assume} that a disease (virus, power blackout) spreads
with a constant radial velocity $v=1$ from an original infection
site of a network. The infected volume $V(t)$ grows according to
the following nonlinear differential equation:
\begin{equation}
\label{4}\frac{dV(t)}{dt}=1+2pV(t-\delta)-\mu(1+2p)V^2(t-\delta)
\end{equation}
where $\delta$ is the time-delay during the spreading. The
nonlinear friction term consists of two parts: the former comes
from the nonlinear interaction within the regular lattice, while
the latter is the nonlinear interaction coming from the newly
added links, which is dependent on the probability $p$ in the N-W
small-world model \cite{L-C-L:2004}. And $\mu>0$ is a measure of
such nonlinear frictions.\\
\indent{It} should be noted that Eq. (\ref{4}) contains as special
cases all the existing linear models of Newman and Watts
\cite{N-W:1999} and Moukarzel \cite{Mou:1999}. When $\mu=0$ and
$\delta=0$, obviously the model produces exponential growth
without limit. Besides, the nonlinear model of Yang
\cite{Yang:2001} hides the effect of the probability $p$ on
bifurcation behaviors in spreading, while our model (\ref{4}) is
more general and uncovers this effect. Therefore, model (\ref{4})
is more suitable for studying the spreading phenomenon in
small-world networks.

Denote $s(t)=V(t)-V^*$, where $V^*$ is the equilibrium of Eq.
(\ref{4}) with $V(t)>0$ and is given by
\begin{equation}
\label{6}V^*=\frac{p+\sqrt{p^2+\mu(1+2p)}}{\mu(1+2p)}
\end{equation}

\indent{To} study the stability of the spreading behaviors in
model (\ref{4}), we select $\mu$ as the parameter of bifurcation
for analysis. We have the following theoretical results for the
existence of Hopf bifurcation in model (\ref{4}), including the
directions, stabilities and periods of the bifurcating periodic
solutions.

\newtheorem{theorem}{Theorem}
\begin{theorem} \label{theorem2} If $\delta<\frac{\pi}{4p}$, then when
the positive parameter $\mu$ passes through the critical value
$\mu_0=\frac{\pi^2-16\delta^2p^2}{16\delta^2(1+2p)}$, there is a
Hopf bifurcation in model (\ref{4}) at its equilibrium $V^*$.
\end{theorem}
{\bf Proof:} See reference \cite{L-C-L:2004}.\\

To state the next theorem, define
\begin{equation*}
\overline{B}=\frac{1}{1-2\sqrt{p^2+\mu(1+2p)}\delta
e^{-i\omega_0\delta}},\qquad g_{20}=2\mu(1+2p)\overline{B},
\end{equation*}
\begin{equation*}
g_{11}=-2\mu(1+2p)\overline{B},\qquad\qquad\qquad
g_{02}=2\mu(1+2p)\overline{B},
\end{equation*}
\begin{equation*}
g_{21}=-2\mu(1+2p)\overline{B}i\left[-2\frac{g_{11}+\overline{g}_{11}
+2\mu(1+2p)}{2\sqrt{p^2+\mu(1+2p)}}+\frac{-g_{20}-\overline{g}_{02}
+2\mu(1+2p)}{2i\omega_0+2\sqrt{p^2+\mu(1+2p)}}\right],
\end{equation*}
\begin{equation*}
C_1(0)=\frac{i}{2\omega_0}\left[g_{20}g_{11}-2\left|g_{11}\right|^2
-\frac{1}{3}\left|g_{02}\right|^2\right]+\frac{g_{21}}{2},
\end{equation*}
\begin{equation*}
\alpha'(0)=\frac{d}{d\mu}\left[\mbox{Re
}\lambda\right]\Big|_{\mu=\mu_0}=\frac{8(1+2p)\delta}{4+\pi^2},\qquad
\omega'(0)=\frac{d}{d\mu}\left[\mbox{Im
}\lambda\right]\Big|_{\mu=\mu_0}=\frac{16\delta(1+2p)}{\pi(4+\pi^2)},
\end{equation*}
\begin{equation*}
\mu_2=-\frac{\mbox{Re }C_1(0)}{\alpha'(0)},\qquad
\tau_2=\frac{-\mbox{Im }C_1(0)+\mu_2\omega'(0)}{\omega_0},\qquad
\beta_2=2\mbox{Re }C_1(0).
\end{equation*}

\begin{theorem}\label{theorem3} Parameter $\mu_2$ determines the
direction of the Hopf bifurcation: If $\mu_2>0\ (<0)$, the Hopf
bifurcation is supercritical (subcritical) and bifurcating
periodic solutions exist for $\mu>\mu_0\ (<\mu_0)$; parameter
$\beta_2$ determines the stability of the bifurcating periodic
solutions: the solutions are orbitally stable (unstable) if
$\beta_2<0\ (>0)$; and parameter $\tau_2$ determines the period of
the bifurcating periodic solutions: the period increases
(decreases) if $\tau_2>0\ (<0)$.
\end{theorem}
{\bf Proof:} See reference \cite{L-C-L:2004}.

\section{Control of bifurcating behaviors in small-world networks}

As can be seen from Theorems \ref{theorem2}-\ref{theorem3}, the
critical value of $\mu_0$ for the occurrence of Hopf bifurcation,
and those values of $\mu_2,\ \tau_2,\ \beta_2$ for verifying the
stability of bifurcating periodic solutions, are all dependent on
the probability $p$, which will be denoted by $\mu_0(p),\
\mu_2(p),\ \tau_2(p),\ \beta_2(p)$ below.

\begin{figure}
\centering
\includegraphics[width=10cm]{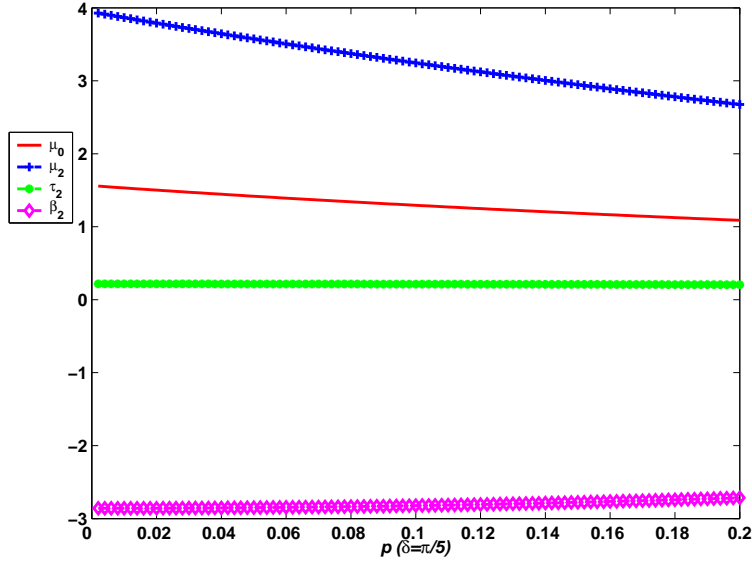}
\caption{\label{fig5} The evolution trend of bifurcation
parameters $\mu_0(p)$, $\mu_2(p)$, $\tau_2(p)$, $\beta_2(p)$, with
$0<p\le 0.2$ in model (\ref{4}) (small-world networks), where
$\delta=\frac{\pi}{5}$.}
\end{figure}

Generally, small-world networks are described by W-S/N-W models
\cite{N-W:1999, W-S:1998} with a small probability $0<p\ll 1$.
Figure \ref{fig5} shows the values of $\mu_0(p)$, $\mu_2(p)$,
$\tau_2(p)$, $\beta_2(p)$ when $0<p\le 0.2$, where, interestingly,
we have found that $\tau_2(p)$ and $\beta_2(p)$ are not sensitive
to the change of the probability $p$. It implies that these
small-world networks have almost the same bifurcating periodic
behaviors, for example, $\tau_2(0.01)=0.2162$, which is almost the
same as $\tau_2(0.1)=0.2135$. This means that when $\mu$ increases
after passing through the critical value $\mu_0(p=0.01,0.1)$, the
periods of these periodic solutions increase in about the same
amount.

Recall that bifurcation control deals with a modification of some
bifurcation characteristics of a parameterized nonlinear system by
a judiciously designed control input \cite{Chen:1999}. Because $p$
determines the bifurcating values of $\mu_0(p)$, $\mu_2(p)$, which
are monotonically decreasing as $p$ increases with a fixed
time-delay $\delta$. Therefore, if we want to apply a controller
to the bifurcating behaviors, $p$ is preferred to be controlled.
We thus employ the following control strategy:
\begin{equation}
\label{5}\frac{dV(t)}{dt}=1+2u(p)V(t-\delta)-\mu(1+2u(p))V^2(t-\delta),
\end{equation}
where
\begin{equation}
\label{7} u(p)=p-\Delta p.
\end{equation}
We may select $p$ as the control parameter to first calculate the
bifurcation parameter $\mu_0(p)$, and then to vary $p$ with
$\Delta p$ so as to obtain a different $\mu_0(p)$. This results in
different bifurcating spreading behaviors in a small-world
network. To this end, we can further stabilize periodic spreading
solutions onto desired stable equilibria, and vice versa.

\begin{figure}
\centering
\includegraphics[width=15cm]{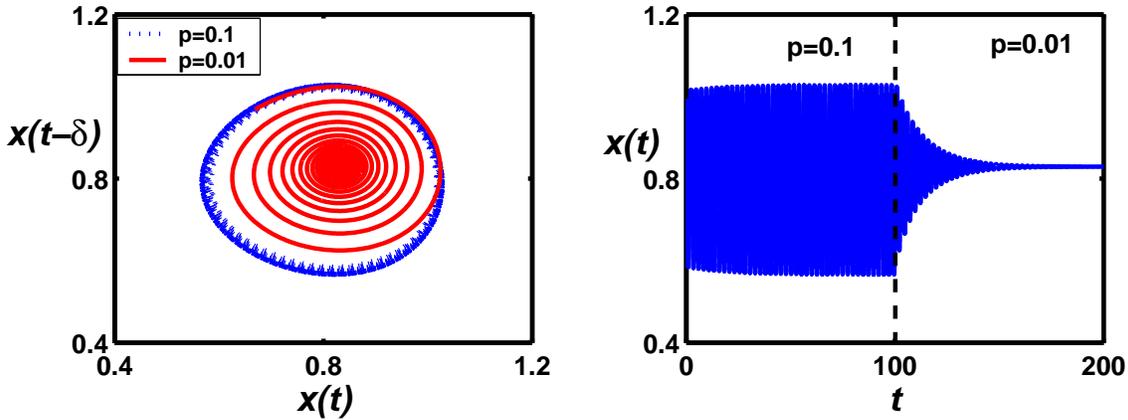}
\caption{\label{fig7} Phase plot and waveform plot of a
small-world network of model (\ref{4}), with $p=0.1$ (dotted
line), and $p=0.01$ (solid line), both $\delta=\frac{\pi}{5}$ and
$\mu=1.45$ ($p$ changes at $t=100$).}
\end{figure}

Figure \ref{fig7} shows an example. At the beginning, $t\in [0,
100]$, we set $p=0.1$, $\delta=\frac{\pi}{5}$, $\mu=1.45$ in model
(\ref{4}), and then compute $\mu_0(0.1)=1.2938<\mu=1.45$. As
expected, a Hopf bifurcation occurs and the spreading behavior is
a periodic solution. To stabilize this periodic orbit onto a
stable equilibrium, we vary the probability $p$ to be 0.01 after
$t=100$. Since $\mu_0(0.01)=1.5315>\mu=1.45$, the periodic
spreading solution asymptotically converges to the stable
equilibrium $V^*=0.8291$, as shown in Fig. \ref{fig7}.

\section{Conclusion and discussion}

In this paper, we have introduced a general nonlinear spreading
model of the small-world network type, which has a flexible
nonlinear interaction effect. Based on the study of Hopf
bifurcation in this spreading model, a simple bifurcation control
example has been simulated, showing that the probability $p$ can
be used as a control parameter to stabilize a periodic spreading
behavior onto a stable equilibrium in the small-world network.\\
More recently, Motter \cite{Motter: 2004} studied a means of
controlling cascade failure in a complex network, and pointed out
that selective removal of network elements can prevent the cascade
from propagating through the entire network. Such a selective
removal control strategy is actually a complementary approach of
the technique discussed in this paper, for controlling the
spreading in small-world networks.

\end{document}